\documentstyle[amssymb,aps,12pt]{revtex}

\begin{document}
\author{J.P. Krisch and E.N. Glass\thanks{%
Permanent address: Physics Department, University of Windsor, Ontario N9B
3P4, Canada}}
\title{A Spacetime in Toroidal Coordinates}
\address{Department of Physics, University of Michigan, Ann Arbor, Michgan 48109}
\date{23 March 2003}
\maketitle

\begin{abstract}
\ \newline
We present an exact solution of Einstein's field equations in toroidal
coordinates. The solution has three regions: an interior with a string
equation of state; an Israel boundary layer; an exterior with constant
isotropic pressure and constant density, locally isometric to anti-de Sitter
spacetime. The exterior can be a cosmological vacuum with negative
cosmological constant.\ The size and mass of the toroidal loop depend on the
size of $\Lambda $. \newline
\newpage
\end{abstract}

\section{Introduction}

There has been increasing interest in spacetimes with non spherical topology
and negative cosmological constant. Much of the discussion has focused on
structures with horizons in anti-de Sitter (AdS) spacetimes \cite{ABH+96}
\cite{SM97}\cite{Lem98}\cite{DL01}\cite{BLP97}. Vanzo \cite{Van97} pointed
out that, in AdS space, there are black hole solutions with genus $g$
horizons. Aminneborg et al (ABHP) \cite{ABH+96} discussed spacetimes locally
isometric to AdS with horizons of arbitrary genus. While many current models
of the universe seem to indicate that $\Lambda $ is positive, there are some
models with $\Lambda <0$ \cite{NVB02}. Aside from their physical relevance
to the actual structure of the Universe, solutions in AdS are very
interesting as a comparison case to asymptotically flat solutions. For
example, Hawking and Page \cite{HP83} have discussed the relevance of a
negative $\Lambda $ to black hole thermal stability. The 2+1
Ba\~{n}ados-Teitelboim-Zanelli \cite{BTZ92} black hole solution and its 3+1
black string \cite{BTZ94} lift have generated a large literature \cite{Pee00}%
.

In this work we discuss a toroidal fluid solution embedded in a locally AdS
exterior. There is an overall metric scale factor which depends on the size
of $\Lambda $. The solution has three regions:\newline
(i) an interior solution with an equation of state, $\rho +p_{\varphi }=0;$%
\newline
(ii) an Israel \cite{Isr77} boundary layer with surface stress energy $%
S_{ij} $ and string-like content $S_{00}+S_{\varphi \varphi }=0;$\newline
(iii) an exterior with constant isotropic pressure, constant density, and a
negative cosmological constant. Just as in the ABHP study, the exterior
metric is locally isometric to AdS. The solution models an extended loop
with interior structure. The size of the loop and its mass depend on the
cosmological constant. The solution can be used to model both micro loops or
very large loop structures, depending on the size of $\Lambda $.

There have been other discussions of circular string structures. Frolov,
Israel and Unruh \cite{FIU89} started with an axially symmetric spacetime
and discussed the relation between internal string structure and angular
deficit, then transformed the metric to toroidal coordinates to discuss the
mass structure of circular cosmic strings. Using toroidal coordinates,
Hughes et al \cite{HMV93} studied weak field loops. Sen and Banerjee \cite
{SB98} have discussed a solution for a circular cosmic string loop in
cylindrical coordinates. Because often a particular choice of surfaces can
simplify the solution of the field equations, we begin with toroidal
coordinates.

Cartesian toroids are discussed in the next section. In section III we write
the field equations for the spacetime and develop the interior and exterior
solutions. Matching conditions are presented in the fourth section. The
Israel boundary layer is described in the fifth section. In section VI we
discuss the mass, and the final section concludes with a general discussion.

\section{Cartesian Toroids}

The relation between Cartesian coordinates $(x,y,z)$ and toroidal
coordinates ($\eta ,\vartheta ,\varphi )$ on ${\cal R}^{3}$ is \cite{MO54}
\begin{mathletters}
\label{xyz-coord}
\begin{eqnarray}
x &=&a\frac{\sinh (\eta )\cos (\varphi )}{\cosh (\eta )-\cos (\vartheta )},
\label{x-coord} \\
y &=&a\frac{\sinh (\eta )\sin (\varphi )}{\cosh (\eta )-\cos (\vartheta )},
\label{y-coord} \\
z &=&a\frac{\sin (\vartheta )}{\cosh (\eta )-\cos (\vartheta )},
\label{z-coord}
\end{eqnarray}
with $0\leq \eta <\infty ,\ 0\leq \vartheta \leq 2\pi ,\ 0\leq \varphi \leq
2\pi $. '$a$' is a constant scale factor.

In toroidal coordinates, the Euclidean metric $dx^{2}+dy^{2}+dz^{2}$ becomes
\end{mathletters}
\begin{equation}
dL^{2}=\frac{a^{2}}{[\cosh (\eta )-\cos (\vartheta )]^{2}}[d\eta
^{2}+d\vartheta ^{2}+\sinh ^{2}(\eta )d\varphi ^{2}]  \label{tor-3met}
\end{equation}
The torus $\eta =\eta _{0}$ described by $dL^{2}$ has a circular cross
section with circumference $2\pi a\csc $h$(\eta _{0})$, and the center of
the toroid circular cross section a distance $a\cot $h$(\eta _{0})$ from the
origin. The equation of the $\varphi =0$, $y=0$ circles, Eq.(\ref{y-coord}),
is \cite{Arf70}
\[
\lbrack x-a\coth (\eta _{0})]^{2}+z^{2}=a^{2}\text{csch}^{2}(\eta _{0})
\]
As $\eta _{0}$ increases, the radius of the loop decreases and the torus
approaches the flat torus, a distance '$a$' from the origin. Looking down
the z-axis (about which $\varphi $ has range $0\leq \varphi \leq 2\pi $) at
the torus, one sees two concentric circles. The $\vartheta =$ constant
surfaces, $0\leq \vartheta \leq 2\pi ,$ are spheres centered on the z-axis.
From Eq.(\ref{xyz-coord}) these spheres have equation
\[
(x^{2}+y^{2}+z^{2}-a^{2})/2az=\cot (\vartheta )
\]
which defines the relation of $\vartheta $ to the torus.

\section{Spacetime}

For the curved space torus, one must construct two different metrics, an
exterior for $0\leq \eta \leq \eta _{0}$ and an interior for $\eta _{0}\leq
\eta \leq \infty $. The metric that we use to describe the spacetime is a
simple generalization of the flat space metric:
\begin{equation}
ds^{2}=\frac{a^{2}}{[\cosh (\eta )-\cos (\vartheta )]^{2}}[-h^{2}(\eta
)dt^{2}+e^{2\mu (\eta )}d\eta ^{2}+d\vartheta ^{2}+h^{2}(\eta )d\varphi
^{2}].  \label{gen-met}
\end{equation}
Note that metric (\ref{gen-met}) cannot reduce to the Minkowski metric.

\subsection{Field Equations}

We write Einstein's field equations using the conventions of Misner, Thorne,
and Wheeler \cite{MTW73} and Wald \cite{Wal84}. The field equations are ($%
G=c=1$)
\begin{equation}
G_{\alpha \beta }=8\pi T_{\alpha \beta }.
\end{equation}
Using flow vector $\hat{u}^{\alpha }\hat{u}_{\alpha }=-1$, the
energy-momentum tensor for a fluid is given in terms of principal pressures
as
\begin{equation}
T_{\alpha \beta }=\rho \hat{u}_{\alpha }\hat{u}_{\beta }+p_{1}\hat{x}%
_{\alpha }\hat{x}_{\beta }+p_{2}\hat{y}_{\alpha }\hat{y}_{\beta }+p_{3}\hat{z%
}_{\alpha }\hat{z}_{\beta }.
\end{equation}

In the following development, we write the field equations allowing for
fluid anisotropy. We do not include $\Lambda $ explicitly in the
stress-energy tensor but will interpret the stress-energy associated with a
metric solution in terms of $\Lambda $ if appropriate. Using metric (\ref
{gen-met}) above with $\Psi =\cosh (\eta )-\cos (\vartheta )$ and $u^{\alpha
}\partial _{\alpha }=(\Psi /ah)\partial _{t}$, the field equations are
\begin{mathletters}
\label{set1}
\begin{eqnarray}
8\pi \rho a^{2}e^{2\mu } &=&-8\pi p_{\varphi }a^{2}e^{2\mu }=-\cosh
^{2}(\eta )-2\cosh (\eta )\cos (\vartheta )+3+2\sinh (\eta )\Psi (h^{\prime
}/h)  \label{rho-eqn} \\
&&-\Psi ^{2}(h^{\prime \prime }/h)+\Psi ^{2}\mu ^{\prime }(h^{\prime
}/h)-2\Psi \mu ^{\prime }\sinh (\eta )+e^{2\mu }[-3+\cos ^{2}(\vartheta
)+2\cosh (\eta )\cos (\vartheta )]  \nonumber \\
8\pi p_{\eta }a^{2}e^{2\mu } &=&3\cosh ^{2}(\eta )-3-4\Psi \sinh (\eta
)(h^{\prime }/h)+\Psi ^{2}(h^{\prime }/h)^{2}  \label{p-eta-eqn} \\
&&+e^{2\mu }[3-2\cosh (\eta )\cos (\vartheta )-\cos ^{2}(\vartheta )]
\nonumber \\
8\pi p_{\vartheta }a^{2}e^{2\mu } &=&\cosh ^{2}(\eta )-3+2\cosh (\eta )\cos
(\vartheta )-2\Psi \sinh (\eta )[2(h^{\prime }/h)-\mu ^{\prime }]
\label{p-theta-eqn} \\
&&+\Psi ^{2}[2(h^{\prime \prime }/h)-2\mu ^{\prime }(h^{\prime
}/h)+(h^{\prime }/h)^{2}]+3e^{2\mu }\sin ^{2}(\vartheta )  \nonumber
\end{eqnarray}
where $\partial h/\partial \eta $ and $\partial \mu /\partial \eta $ are
abbreviated by $h^{\prime }$ and $\mu ^{\prime }$.

\subsection{Interior solution}

Let $h^{2}=[d_{0}$sinh$(\eta )-b_{0}]^{2},\ e^{2\mu }=1$. The interior
metric is
\end{mathletters}
\begin{equation}
g_{\alpha \beta }^{\text{in}}dx^{\alpha }dx^{\beta }=(a/\Psi
)^{2}[-h^{2}dt^{2}+d\eta ^{2}+d\vartheta ^{2}+h^{2}d\varphi ^{2}].
\label{in-met}
\end{equation}
The energy-momentum components for $g^{\text{in}}$ are
\begin{mathletters}
\label{set2}
\begin{eqnarray}
8\pi a^{2}\rho &=&-8\pi a^{2}p_{\varphi }=(b_{0}/h)[\cosh ^{2}(\eta )-\cos
^{2}(\vartheta )],  \label{rho-p-phi-in} \\
8\pi a^{2}p_{\eta } &=&(\Psi /h^{2})\{(d_{0}^{2}+b_{0}^{2})\Psi
-2b_{0}h[\cosh (\eta )+\cos (\vartheta )]\},  \label{p-eta-in} \\
8\pi a^{2}p_{\vartheta } &=&(\Psi /h^{2})[(d_{0}^{2}+b_{0}^{2})\Psi
-4b_{0}h\cos (\vartheta )].  \label{p-thet-in}
\end{eqnarray}
The equation of state is
\end{mathletters}
\begin{equation}
\rho +p_{\varphi }=0.
\end{equation}
The interior metric has quadratic Weyl invariant
\begin{equation}
C_{\alpha \beta \mu \nu }C^{\alpha \beta \mu \nu }=\frac{4}{3}d_{0}^{2}\frac{%
\Psi ^{4}}{a^{4}h^{4}}[b_{0}\sinh (\eta )+d_{0}]^{2},
\end{equation}
and Ricci scalar
\begin{equation}
R_{\alpha \beta }g_{\text{in}}^{\alpha \beta }=-\frac{2\Psi }{a^{2}h^{2}}%
\left\{ (d_{0}^{2}+b_{0}^{2})\Psi -2b_{0}h[\cosh (\eta )+2\cos (\vartheta
)]\right\} {\huge .}
\end{equation}

\subsection{Exterior Solution}

The solution to be used in the toroid exterior is
\begin{equation}
g_{\alpha \beta }^{\text{ex}}dx^{\alpha }dx^{\beta }=\frac{a^{2}}{[\cosh
(\eta )-\cos (\vartheta )]^{2}}[-h^{2}(\eta )dt^{2}+e^{2\mu (\eta )}d\eta
^{2}+d\vartheta ^{2}+h^{2}(\eta )d\varphi ^{2}].  \label{ex-met}
\end{equation}
In order to describe a cosmological vacuum, $p_{\eta }$ will have to be
constant . The cosine terms should vanish. From the general field equations
we write $p_{\eta }$, grouping the terms:
\begin{eqnarray*}
8\pi p_{\eta }a^{2}e^{2\mu } &=&3\cosh ^{2}(\eta )-3-4\sinh (\eta )\cosh
(\eta )(h^{\prime }/h)+3e^{2\mu }+\cosh ^{2}(\eta )(h^{\prime }/h)^{2} \\
&&+\cos (\vartheta )[-2\cosh (\eta )e^{2\mu }+4\sinh (\eta )(h^{\prime
}/h)-2\cosh (\eta )(h^{\prime }/h)^{2}] \\
&&+\cos ^{2}(\vartheta )[-e^{2\mu }+(h^{\prime }/h)^{2}]
\end{eqnarray*}
To eliminate the $\cos ^{2}$ term, take$\ (h^{\prime }/h)^{2}=e^{2\mu }$.
The cosine term then becomes
\[
4\cos (\vartheta )e^{\mu }[-\cosh (\eta )e^{\mu }+\sinh (\eta )].
\]
Requiring this term to vanish provides one non-trivial solution
\begin{equation}
e^{\mu }=\sinh (\eta )/\cosh (\eta ),\text{ \ }h=\cosh (\eta ).
\label{soln-a}
\end{equation}
Substituting (\ref{soln-a}), the energy-momentum components of $g^{\text{ex}}
$ are
\begin{mathletters}
\label{set3}
\begin{eqnarray}
8\pi \rho  &=&-3/a^{2},  \label{rho-stress} \\
8\pi p_{\eta } &=&8\pi p_{\vartheta }=8\pi p_{\varphi }=3/a^{2}.
\label{phi-thet-eta-stress}
\end{eqnarray}
This can be a spacetime with negative cosmological constant $\Lambda
=-3/a^{2}$. The metric is conformally flat and has constant negative Ricci
scalar $R=-12/a^{2}$. $g_{\text{ex}}^{\alpha \beta }$ is locally isometric
to the AdS metric.

\section{Matching Interior to Exterior}

The two metrics to be joined are
\end{mathletters}
\begin{eqnarray}
g_{\alpha \beta }^{\text{in}}dx^{\alpha }dx^{\beta } &=&\frac{a^{2}}{\Psi
^{2}}\{-[d_{0}\text{sinh}(\eta )-b_{0}]^{2}dt^{2}+d\eta ^{2}+d\vartheta
^{2}+[d_{0}\text{sinh}(\eta )-b_{0}]^{2}d\varphi ^{2}\} \\
g_{\alpha \beta }^{\text{ex}}dx^{\alpha }dx^{\beta } &=&\frac{a^{2}}{\Psi
^{2}}[-\cosh ^{2}(\eta )dt^{2}+\frac{\sinh ^{2}(\eta )}{\cosh ^{2}(\eta )}%
d\eta ^{2}+d\vartheta ^{2}+\cosh ^{2}(\eta )d\varphi ^{2}]  \nonumber
\end{eqnarray}
Matching the metrics one obtains
\[
\cosh (\eta _{0})=d_{0}\sinh (\eta _{0})-b_{0}\text{.}
\]
Matching the extrinsic curvature yields
\[
d_{0}\cosh (\eta _{0})=\sinh (\eta _{0}).
\]
The bounding surface is thus defined by
\begin{mathletters}
\label{first-bound}
\begin{eqnarray}
\cosh (\eta _{0}) &=&\frac{b_{0}}{d_{0}^{2}-1}  \label{cosh} \\
\sinh (\eta _{0}) &=&\frac{d_{0}b_{0}}{d_{0}^{2}-1}  \label{sinh}
\end{eqnarray}
with
\end{mathletters}
\begin{equation}
b_{0}^{2}+d_{0}^{2}=1
\end{equation}
This implies that both $b_{0}$ and $d_{0}$ are less than $1.$ On the
boundary the stresses are
\begin{mathletters}
\label{set4}
\begin{eqnarray}
8\pi a^{2}\rho &=&-8\pi a^{2}p_{\varphi }=[b_{0}^{2}\cos ^{2}(\vartheta )-1],
\label{rho-p-match} \\
8\pi a^{2}p_{\eta } &=&\Psi b_{0}[-3+b_{0}\cos (\vartheta )],
\label{p-eta-match} \\
8\pi a^{2}p_{\vartheta } &=&\Psi b_{0}[-1+3b_{0}\cos (\vartheta )].
\label{p-thet-match}
\end{eqnarray}
A problem with the matching is that $p_{\eta }$ does not smoothly join to
the exterior stress. This mismatch would lead to a dynamic boundary.
Therefore, an Israel boundary layer will be developed.

\section{The Boundary Layer}

\subsection{Position of the Layer}

If the interior and exterior solutions do not match derivatives but joined
over an Israel surface layer \cite{Isr77}, then the position of the boundary
will be set by matching only $h$ at $\eta =\eta _{0}.$ For the exterior we
have
\end{mathletters}
\[
h=\text{cosh}(\eta ),\ \ e^{\mu }=\sinh (\eta )/\cosh (\eta ).
\]
For the interior
\[
h=d_{0}\text{sinh}(\eta )-b_{0},\ \text{\ }e^{2\mu }=1.
\]
Matching the interior and exterior at $\eta =\eta _{0}$ provides
\[
\cosh (\eta _{0})=d_{0}\sinh (\eta _{0})-b_{0}\text{.}
\]
Note that the $e^{2\mu }$ term need not match, since it is the coefficient
of $d\eta ^{2}$ and the match is for $\eta $ constant surfaces. Rearranging,
we have the bounding surface
\begin{mathletters}
\label{bound-val}
\begin{eqnarray}
\cosh (\eta _{0}) &=&\frac{b_{0}+\Bbbk d_{0}(b_{0}^{2}+d_{0}^{2}-1)^{1/2}}{%
d_{0}^{2}-1},\text{ \ }\Bbbk =(\pm 1)  \label{cosh-2} \\
\sinh (\eta _{0}) &=&\frac{d_{0}b_{0}+\Bbbk (b_{0}^{2}+d_{0}^{2}-1)^{1/2}}{%
d_{0}^{2}-1}.  \label{sinh-2}
\end{eqnarray}

\subsection{Parameter Constraints}

Constraints can be set on $d_{0}$ and $b_{0}$ by requiring
\end{mathletters}
\[
\sinh (\eta _{0})>0,\text{ \ }\cosh (\eta _{0})>0,\text{ \ }\rho _{\text{%
interior}}>0.
\]
Both of the hyperbolic functions in Eq.(\ref{bound-val}) have a sign choice
which is the same for both functions. There are eight possible parameter ($%
\Bbbk ,d_{0},b_{0})$ combinations for both $d_{0}^{2}>1$ and $d_{0}^{2}<1$
for a total of sixteen cases. The hyperbolic conditions eliminate eight and
the density constraint five more. The three remaining allowed parameter
combinations with their constraints are:
\begin{eqnarray}
\text{(1)}\ d_{0}^{2} &>&1:[\Bbbk =+1,\ d_{0}>0,\ b_{0}>0],\text{ no
constraints}  \label{3constraints} \\
\text{(2)}\ d_{0}^{2} &>&1:[\Bbbk =-1,\ d_{0}>0,\ b_{0}>0],\ \sqrt{%
b_{0}^{2}+d_{0}^{2}-1}<\left| \frac{b_{0}}{d_{0}}\right| ,\ \sqrt{%
b_{0}^{2}+d_{0}^{2}-1}<\left| d_{0}b_{0}\right|  \nonumber \\
\text{(3)\ }d_{0}^{2} &<&1:[\Bbbk =-1,\ d_{0}>0,\ b_{0}>0],\ \sqrt{%
b_{0}^{2}+d_{0}^{2}-1}>\left| \frac{b_{0}}{d_{0}}\right| ,\ \sqrt{%
b_{0}^{2}+d_{0}^{2}-1}>\left| d_{0}b_{0}\right|  \nonumber
\end{eqnarray}

The algebraic details are in Appendix A.

\subsection{Extrinsic Curvature}

We are interested in a spacetime that could describe a loop of matter with
an energy density equal to the loop tension over a bounding Israel surface
layer at $\eta =\eta _{0}$. The stress-energy content of the surface layer $%
S_{ij}$ \cite{Isr77} is given by
\begin{equation}
8\pi S_{ij}=\gamma _{ij}-\gamma h_{ij}^{(b)}
\end{equation}
with $h_{ij}^{(b)}$ the metric of the bounding torus. $\gamma _{ij}$ is the
difference between the extrinsic curvatures of the exterior and interior
metrics on the boundary
\[
\gamma _{ij}=K_{ij}^{\text{ex}}-K_{ij}^{\text{in}}=\ <K_{ij}>.
\]
Calculating the general extrinsic curvature on the bounding torus $\eta
=\eta _{0}$ with unit normal $n_{\alpha }$ we have
\begin{eqnarray*}
K_{ij} &=&-n_{\alpha ;\beta }h_{i}^{\alpha }h_{j}^{\beta } \\
K_{ij} &=&n_{\alpha }\Gamma _{ij}^{\alpha }=-(n_{\alpha }/2)g^{\alpha \beta
}g_{ij,\beta }
\end{eqnarray*}
With $\Psi =$\ cosh$(\eta )-\cos (\vartheta )$ and $\eta ^{\alpha }\partial
_{\alpha }=\partial /\partial \eta $, we have for the extrinsic curvatures
on the boundary
\begin{mathletters}
\label{many-k}
\begin{eqnarray}
K_{00} &=&(n_{\alpha }\eta ^{\alpha })\frac{\Psi ^{2}}{2e^{2\mu }}\frac{%
\partial }{\partial \eta }(h^{2}/\Psi ^{2}),  \label{k-0} \\
K_{\varphi \varphi } &=&-K_{00},  \label{k-phi} \\
K_{\vartheta \vartheta } &=&(n_{\alpha }\eta ^{\alpha })\frac{\Psi ^{2}}{%
2e^{2\mu }}\frac{\partial }{\partial \eta }(1/\Psi ^{2})  \label{k-thet}
\end{eqnarray}
$K_{\vartheta \vartheta \text{ }}$ will match across the boundary with the
metrics we have found. Using equation (\ref{k-0}) and forming $K_{00}$ we
have
\end{mathletters}
\begin{equation}
K_{00}=\frac{h}{\Psi }[\Psi h^{\prime }-h\sinh (\eta )]  \nonumber
\end{equation}
Establishing the difference between inner and outer spaces and matching $h$
on the boundary, the discontinuity in the extrinsic curvature is
\[
<K_{00}>\ =h[\sinh (\eta _{0})-d_{0}\cosh (\eta _{0})].
\]
Therefore the boundary layer has a stress energy content
\begin{equation}
8\pi S_{00}=\cosh (\eta _{0})[d_{0}\cosh (\eta _{0})-\sinh (\eta
_{0})]=-8\pi S_{\varphi \varphi }.  \label{s-content}
\end{equation}
Requiring $S_{00}>0$ and substituting for $\cosh (\eta _{0})$ and $\sinh
(\eta _{0})$ from Eq.(\ref{bound-val}) implies $\Bbbk =1$. Thus one
parameter set remains:
\begin{equation}
d_{0}^{2}>1:[\Bbbk =+1,\ d_{0}>0,\ b_{0}>0]\text{.}  \label{last-set}
\end{equation}
The stress energy content of the boundary layer represents a toroidal loop
with a string-like equation of state.

\section{Mass}

When the generator of time translations is Killing vector $\xi ^{\nu }$ then
the Einstein four-momentum $p^{\mu }=\sqrt{-g}T_{\ \nu }^{\mu }\xi ^{\nu }$
is conserved and a mass can be associated with three-volume $dV_{\mu }$
\[
M=\int\limits_{3vol}\sqrt{-g}T_{\ \nu }^{\mu }\xi ^{\nu }dV_{\mu }
\]
where $dV_{\mu }=t,_{\mu }d\eta d\vartheta d\varphi $. Substituting we have
the mass inside the torus
\begin{eqnarray}
M &=&\frac{2\pi b_{0}a^{2}}{8\pi }\int\limits_{\eta _{0}}^{\infty
}\int\limits_{0}^{2\pi }\frac{h}{\Psi ^{3}}[\cosh (\eta )+\cos (\vartheta
)]d\vartheta d\eta  \label{mass-int} \\
&=&\frac{\pi b_{0}a^{2}}{8\sinh ^{4}(\eta _{0})}\left\{ 4d_{0}\sinh (\eta
_{0})\cosh ^{2}(\eta _{0})-b_{0}[2\sinh ^{2}(\eta _{0})+3]\right\} .
\nonumber
\end{eqnarray}

A similar calculation can be repeated for the mass associated with the
surface layer. In the Israel formalism the surface stress energy is defined
in geodesic coordinates as the thickness $\varepsilon $ of the layer
approaches zero
\begin{equation}
S_{\mu \nu }=\ \stackrel{\text{lim}}{\varepsilon \rightarrow 0}%
\int\limits_{0}^{\varepsilon }T_{\mu \nu }dx
\end{equation}
Start with the definition of the mass in a three-volume and take the limit
as the distance between tori goes to zero.
\begin{eqnarray*}
M^{\prime } &=&\int\limits_{3vol}\sqrt{-g}T_{\ \nu }^{\mu }\xi ^{\nu
}dV_{\mu } \\
&=&\int\limits_{3vol}\sqrt{-g}T_{\ \nu }^{0}\xi ^{\nu }d\eta d\vartheta
d\varphi
\end{eqnarray*}
In the limit of zero layer thickness
\begin{equation}
M^{\prime }=\ \stackrel{\text{lim}}{\varepsilon \rightarrow 0}\int \int
\int\limits_{\eta _{0}-\varepsilon }^{\eta _{0}}\sqrt{-g_{tt}g_{\vartheta
\vartheta }g_{\varphi \varphi }g_{\eta \eta }}\ T_{\ \nu }^{0}\xi ^{\nu
}d\eta d\vartheta d\varphi  \nonumber
\end{equation}
Assume that the limit can be taken inside the integral and that over the
range of the $\eta -$ integral that $\sqrt{-g_{tt}g_{\vartheta \vartheta
}g_{\varphi \varphi }\ }$ is approximately constant and takes its value on $%
\eta _{0}$.
\begin{eqnarray*}
M^{\prime } &=&\int \int \sqrt{-g_{tt}(\eta _{0},\vartheta )g_{\vartheta
\vartheta }(\eta _{0},\vartheta )g_{\varphi \varphi }(\eta _{0},\vartheta )}%
\ d\vartheta d\varphi \ \stackrel{\text{lim}}{\varepsilon \rightarrow 0}%
\int\limits_{\eta _{0}-\varepsilon }^{\eta _{0}}(T_{\ \nu }^{0}\xi ^{\nu }%
\sqrt{g_{\eta \eta }}d\eta ) \\
M^{\prime } &=&\int \int \sqrt{-g_{tt}g_{\vartheta \vartheta }g_{\varphi
\varphi }}\ S_{\ \nu }^{0}\xi ^{\nu }d\vartheta d\varphi
\end{eqnarray*}
Integration results in
\begin{equation}
M^{\prime }=\frac{ah^{2}}{4}[d_{0}-\tanh (\eta _{0})]\frac{2\pi }{\sinh
(\eta _{0})}.  \label{m-prime-layer}
\end{equation}

\section{Discussion}

In summary, we have obtained a fluid solution to the field equations that
describes a positive density torus with a boundary layer, embedded in a
locally AdS exterior. The solution has two parameters, $d_{0}$ and $b_{0}$
with a restricted range. The fluid and boundary layer both have a
string-like equation of state. The solution can describe a variety of
structures, depending on the parameter value chosen. First consider the size
of the loop, $R_{\vartheta }=a\csc $h$(\eta _{0}).$ For the allowed
parameter set we have, in the limit $b_{0}^{2}>>\left| d_{0}^{2}-1\right| $,
\[
\frac{R_{\vartheta }[\Bbbk =+1,\ d_{0}^{2}>1]}{a}\sim \frac{d_{0}-1}{b_{0}}.
\]
$R_{\vartheta }/a$ can become very small and the torus will approach the
flat torus a distance '$a$' from the center of the torus loop. The size of
the loop depends on the scale parameter, '$a$'. The size of the scale factor
is determined by the cosmological constant. From the field equations we have
\begin{equation}
\frac{8\pi G}{c^{2}}\rho _{\text{exterior}}=-\frac{3}{a^{2}}\text{, \ \ \ }%
\left| \Lambda \right| =\frac{3}{a^{2}}
\end{equation}
For example, if this density is roughly the same order as the critical
density we would have $\left| \rho \right| \sim 10^{-27}\ $kg/m$^{3}\ $and
one finds that $a\sim 10^{28}$ m. If the solution is used to describe a
primordial universe with a large negative $\Lambda ,$ the scale factor could
be much smaller and micro loops could be possible.

The mass description is also dependent on the size of the scale factor. We
have from Eq.(\ref{mass-int}) for the fluid interior
\[
M=\frac{\pi b_{0}a^{2}}{8}\left[ \frac{4d_{0}}{\sinh (\eta _{0})}+\frac{%
4d_{0}}{\sinh ^{3}(\eta _{0})}-\frac{2b_{0}}{\sinh ^{2}(\eta _{0})}-\frac{%
3b_{0}}{\sinh ^{4}(\eta _{0})}\right] .
\]
For the surface layer we have Eq.(\ref{m-prime-layer})
\[
M^{\prime }=\frac{\pi }{2}ah^{2}[d_{0}-\tanh (\eta _{0})]\frac{1}{\sinh
(\eta _{0})}.
\]
One thing that is immediately obvious is the different dependence on the
scale parameter. In the large $b_{0}$ limit taken above we have
\begin{eqnarray*}
M^{\prime } &\sim &\frac{\pi }{2}ab_{0}, \\
M &\sim &\frac{\pi }{4}a^{2}(d_{0}^{2}-1).
\end{eqnarray*}
The fluid inside the torus does not depend on $b_{0}$ in this limit. In the
current universe, if $a>>1$ and if $b_{0}<<a$, the fluid inside the torus
can dominate the mass because of the scale factor. If $b_{0}\sim a$ and $%
d_{0}\rightarrow 1,$ the mass in the surface layer could dominate the loop
structure. While the size of the thin-loop torus depends on '$a$', the
''fat'' torus can extend much closer in to the origin. As above, if, in the
primordial universe, the cosmological constant was negative and much larger,
the scale factor, '$a$'$,$ could be quite small.\ The solution could then
describe micro loops with the surface layer the dominant mass contribution.

Several extensions of this solution might be possible. Adding time
dependence to generate an oscillating loop for a Casimir calculation would
be quite interesting. Time dependence could also be used to check the
evolution and stability over time of a primordial loop. This solution could
also be regarded as a step toward generating multi segment Brevik-Nielson
\cite{BN90} loops with metric dependent tensions. \appendix

\section{MATCHING CONSTRAINTS}

The hyperbolic functions are, with $%
S(b_{0},d_{0}):=(b_{0}^{2}+d_{0}^{2}-1)^{1/2}$,
\begin{mathletters}
\label{many-k}
\begin{eqnarray}
\cosh (\eta _{0}) &=&\frac{b_{0}+\Bbbk d_{0}S}{d_{0}^{2}-1},\text{ \ }\Bbbk
=(\pm 1)  \label{cosh-2} \\
\sinh (\eta _{0}) &=&\frac{d_{0}b_{0}+\Bbbk S}{d_{0}^{2}-1}.  \label{sinh-2}
\end{eqnarray}
The conditions to be satisfied are
\end{mathletters}
\begin{eqnarray*}
\sinh (\eta _{0}) &>&0, \\
\cosh (\eta _{0}) &>&0.
\end{eqnarray*}
The cosh function is always positive and $\sinh (\eta _{0})$ is positive
because the range for the interior metric is $\eta _{0}<\eta <\infty .$ The
parameters must always satisfy the condition
\[
b_{0}^{2}+d_{0}^{2}>1.
\]
The equal sign with $S=0$ is not a possibility since that would imply an
exact match of interior and exterior.

\subsection{sinh$(\protect\eta _{0})>0$}

\[
\frac{d_{0}b_{0}+\Bbbk S}{d_{0}^{2}-1}>0
\]
A: $d_{0}^{2}>1$, $\Bbbk =+1,$ $0<d_{0}b_{0}+S$

(1) $\ (d_{0}>0,\ b_{0}>0)$ condition satisfied

(2) $\ (d_{0}<0,\ b_{0}<0)$ condition satisfied

(3) $\ (d_{0}>0,\ b_{0}<0)$ condition satisfied if $\left| d_{0}b_{0}\right|
<S$

(4) $\ (d_{0}<0,\ b_{0}>0)$ condition satisfied if $\left| d_{0}b_{0}\right|
<S$\newline
B: $d_{0}^{2}>1$, $\Bbbk =-1,$ $0<d_{0}b_{0}-S$

(5)$\ \ (d_{0}>0,\ b_{0}>0)$ condition satisfied if $S<\left|
d_{0}b_{0}\right| $

(6)\ $\ (d_{0}<0,\ b_{0}<0)$ condition satisfied if $S<\left|
d_{0}b_{0}\right| $

(7) $\ (d_{0}>0,\ b_{0}<0)$ condition excluded

(8) \ $(d_{0}<0,\ b_{0}>0)$ condition excluded\newline
C: $d_{0}^{2}<1$, $\Bbbk =+1,$ $0<-d_{0}b_{0}-S$

(9)\ $\ (d_{0}>0,\ b_{0}>0)$ condition excluded

(10) $(d_{0}<0,\ b_{0}<0)$ condition excluded

(11) $(d_{0}>0,\ b_{0}<0)$\ condition satisfied if $S<\left|
d_{0}b_{0}\right| $

(12) $(d_{0}<0,\ b_{0}>0)$\ condition satisfied if $S<\left|
d_{0}b_{0}\right| $\newline
D: $d_{0}^{2}<1$, $\Bbbk =-1,$ $0<-d_{0}b_{0}+S$

(13) $(d_{0}>0,\ b_{0}>0)$ condition satisfied if $\left| d_{0}b_{0}\right|
<S$

(14) $(d_{0}<0,\ b_{0}<0)$ condition satisfied if $\left| d_{0}b_{0}\right|
<S$

(15) $(d_{0}>0,\ b_{0}<0)$ condition satisfied

(16)$\ (d_{0}<0,\ b_{0}>0)$ condition satisfied

\ \newline
\frame{Summary of Condition 1}

$d_{0}^{2}>1$, $\Bbbk =-1$, $(d_{0}>0,b_{0}<0)$ and $(d_{0}<0,b_{0}>0)$ are
excluded

$d_{0}^{2}<1$, $\Bbbk =+1$, $(d_{0}>0,b_{0}>0)$ and $(d_{0}<0,b_{0}<0)$ are
excluded

\subsection{cosh$(\protect\eta _{0})>0$}

\[
\frac{b_{0}+d_{0}\Bbbk S}{d_{0}^{2}-1}>0
\]
A: $d_{0}^{2}>1$, $\Bbbk =+1,$ $0<b_{0}+d_{0}S$

(1) \ ($d_{0}>0,\ b_{0}>0)$ condition satisfied

(2) \ ($d_{0}<0,\ b_{0}<0)$ condition excluded

(3) \ ($d_{0}>0,\ b_{0}<0)$ condition satisfied if $\left| b_{0}\right|
<d_{0}S$

(4) \ ($d_{0}<0,\ b_{0}>0)$ condition satisfied if $\left| b_{0}\right|
>\left| d_{0}\right| S$\newline
B: $d_{0}^{2}>1$, $\Bbbk =-1,$ $0<b_{0}-d_{0}S$

(5) \ ($d_{0}>0,\ b_{0}>0)\ $condition satisfied if $d_{0}S<b_{0}$

(6)\ $\ (d_{0}<0,\ b_{0}<0)\ $condition satisfied if $\left| d_{0}\right|
S>\left| b_{0}\right| $

(7)\ $\ (d_{0}<0,\ b_{0}>0)$ condition satisfied

(8)\ $\ (d_{0}>0,\ b_{0}<0)$ condition excluded\newline
C: $d_{0}^{2}<1$, $\Bbbk =+1,$ $0<-b_{0}-d_{0}S$

(9) $\ (d_{0}>0,\ b_{0}>0)\ $condition excluded

(10) $(d_{0}<0,\ b_{0}<0)$ condition satisfied

(11) $(d_{0}>0,\ b_{0}<0)\ $condition satisfied if $\left| b_{0}\right|
>d_{0}S$

(12) $(d_{0}<0,\ b_{0}>0)\ $condition satisfied if $b_{0}<\left|
d_{0}\right| S$\newline
D: $d_{0}^{2}<1$, $\Bbbk =-1,$ $0<-b_{0}+d_{0}S$

(13)$\ (d_{0}>0,\ b_{0}>0)\ $condition satisfied if $\left| b_{0}\right|
<d_{0}S$

(14) $(d_{0}<0,\ b_{0}<0)$ condition satisfied if $\left| b_{0}\right|
>\left| d_{0}\right| S$

(15) $(d_{0}<0,\ b_{0}>0)$ condition excluded

(16) $(d_{0}>0,\ b_{0}<0)$ condition satisfied

\ \newline
\frame{Summary of Condition 2}

$d_{0}^{2}>1$, $\Bbbk =+1$, $(d_{0}<0,\ b_{0}<0)$ is excluded

$d_{0}^{2}>1$, $\Bbbk =-1$, $(d_{0}>0,\ b_{0}<0)$ is excluded

$d_{0}^{2}<1$, $\Bbbk =+1$, $(d_{0}>0,\ b_{0}>0)$ is excluded

$d_{0}^{2}<1$, $\Bbbk =-1$, $(d_{0}<0,\ b_{0}>0)$ is excluded

\ \newline
When the constraints for the two conditions are put together, the cases

$k=+1,\ d_{0}<0,\ b_{0}>0,$ are eliminated for both $d_{0}^{2}>1$ and $%
d_{0}^{2}<1.$

\ \newline
\underline{Summary of existing cases after hyperbolic conditions are imposed}
\begin{eqnarray}
d_{0}^{2} &>&1:\Bbbk =+1 \\
(d_{0} &>&0,\ b_{0}>0)  \nonumber \\
(d_{0} &>&0,\ b_{0}<0):\ \left| d_{0}b_{0}\right| <S,\ \left| \frac{b_{0}}{%
d_{0}}\right| <S  \nonumber
\end{eqnarray}
\begin{eqnarray}
d_{0}^{2} &>&1:\Bbbk =-1 \\
(d_{0} &>&0,\ b_{0}>0):\text{ }S<\left| d_{0}b_{0}\right| ,\ S<\left| \frac{%
b_{0}}{d_{0}}\right|  \nonumber \\
(d_{0} &<&0,\ b_{0}<0):\ S<\left| d_{0}b_{0}\right| ,\ S>\left| \frac{b_{0}}{%
d_{0}}\right|  \nonumber
\end{eqnarray}
\begin{eqnarray}
d_{0}^{2} &<&1:\Bbbk =+1 \\
(d_{0} &>&0,\ b_{0}<0):\text{ }S<\left| d_{0}b_{0}\right| ,\ S<\left| \frac{%
b_{0}}{d_{0}}\right|  \nonumber
\end{eqnarray}
\begin{eqnarray}
d_{0}^{2} &<&1:\Bbbk =-1 \\
(d_{0} &>&0,\ b_{0}>0):\ S>\left| d_{0}b_{0}\right| ,\ S>\left| \frac{b_{0}}{%
d_{0}}\right|  \nonumber \\
(d_{0} &<&0,\ b_{0}<0):\ S>\left| d_{0}b_{0}\right| ,\ S<\left| \frac{b_{0}}{%
d_{0}}\right|  \nonumber \\
(d_{0} &>&0,\ b_{0}<0);  \nonumber
\end{eqnarray}

Now we require the fluid density inside the torus to be positive:
\[
8\pi a^{2}\rho =(b_{0}/h)[\cosh ^{2}(\eta _{0})-\cos ^{2}(\vartheta )]>0
\]
cosh($\eta _{0}$) will always be greater than 1 since it equals 1 at $\eta
=0,$ which is outside of the torus interior. In the interior $\eta _{0}\leq
\eta \leq \infty .$ We have
\begin{eqnarray*}
\frac{b_{0}}{d_{0}\sinh (\eta _{0})-b_{0}} &>&0 \\
\frac{1}{\frac{d_{0}}{b_{0}}\sinh (\eta _{0})-1} &>&0 \\
\frac{d_{0}}{b_{0}}\frac{b_{0}d_{0}+\Bbbk S}{d_{0}^{2}-1} &>&1
\end{eqnarray*}

\subsection{$d_{0}^{2}>1$}

\begin{eqnarray*}
d_{0}^{2}+\Bbbk \frac{d_{0}}{b_{0}}S &>&d_{0}^{2}-1 \\
-\Bbbk \frac{d_{0}}{b_{0}}S &<&1
\end{eqnarray*}

\begin{eqnarray}
\Bbbk &=&+1,\ (b_{0}>0,\ d_{0}>0)\text{ and }(b_{0}<0,\ d_{0}<0)\text{. No
constraints}  \nonumber \\
\Bbbk &=&-1,\ (b_{0}>0,\ d_{0}>0)\text{ and }(b_{0}<0,\ d_{0}<0)\text{ with
constraint }\left| \frac{d_{0}}{b_{0}}\right| S<1  \nonumber
\end{eqnarray}

\subsection{$d_{0}^{2}<1$}

\begin{eqnarray*}
-d_{0}^{2}-\Bbbk \frac{d_{0}}{b_{0}}S &>&1-d_{0}^{2} \\
-\Bbbk \frac{d_{0}}{b_{0}}S &>&1
\end{eqnarray*}
\begin{eqnarray}
\Bbbk &=&+1,\ (b_{0}<0,\ d_{0}>0)\text{ with constraint }S>\left| \frac{b_{0}%
}{d_{0}}\right| \\
\Bbbk &=&-1,\ (b_{0}>0,\ d_{0}>0)\text{ and }(b_{0}<0,\ d_{0}<0)\text{ with
constraint }S>\left| \frac{b_{0}}{d_{0}}\right|  \nonumber
\end{eqnarray}
Summarizing all constraints provides
\begin{eqnarray}
d_{0}^{2} &>&1:\Bbbk =+1 \\
(d_{0} &>&0,\ b_{0}>0)  \nonumber \\
(d_{0} &>&0,\ b_{0}<0):\ \left| d_{0}b_{0}\right| <S,\ \left| \frac{b_{0}}{%
d_{0}}\right| <S,\ S<\left| \frac{b_{0}}{d_{0}}\right| \text{ is excluded}
\nonumber
\end{eqnarray}
\begin{eqnarray}
d_{0}^{2} &>&1:\Bbbk =-1 \\
(d_{0} &>&0,\ b_{0}>0):\text{ }S<\left| d_{0}b_{0}\right| ,\ S<\left| \frac{%
b_{0}}{d_{0}}\right|  \nonumber \\
(d_{0} &<&0,\ b_{0}<0):\ S<\left| d_{0}b_{0}\right| ,\ S>\left| \frac{b_{0}}{%
d_{0}}\right| ,\ S<\left| \frac{b_{0}}{d_{0}}\right| \text{ is excluded}
\nonumber
\end{eqnarray}
\begin{eqnarray}
d_{0}^{2} &<&1:\Bbbk =+1 \\
(d_{0} &>&0,\ b_{0}<0):S<\left| d_{0}b_{0}\right| ,\ S<\left| \frac{b_{0}}{%
d_{0}}\right| ,\ S>\left| \frac{b_{0}}{d_{0}}\right| \text{ is excluded}
\nonumber
\end{eqnarray}
\begin{eqnarray}
d_{0}^{2} &<&1:\Bbbk =-1 \\
(d_{0} &>&0,\ b_{0}>0):\ S>\left| d_{0}b_{0}\right| ,\ S>\left| \frac{b_{0}}{%
d_{0}}\right|  \nonumber \\
(d_{0} &<&0,\ b_{0}<0):\ S>\left| d_{0}b_{0}\right| ,\ S<\left| \frac{b_{0}}{%
d_{0}}\right| ,\ S>\left| \frac{b_{0}}{d_{0}}\right| \text{ is excluded}
\nonumber \\
(d_{0} &>&0,b_{0}<0)\text{ is excluded}  \nonumber
\end{eqnarray}
\

\underline{The three allowed parameter combinations are}

$d_{0}^{2}>1:$ $\Bbbk =+1$ ($d_{0}>0,\ b_{0}>0)$

$d_{0}^{2}>1:$ $\Bbbk =-1$ ($d_{0}>0,\ b_{0}>0)$: $S<\left|
d_{0}b_{0}\right| ,\ S<\left| \frac{b_{0}}{d_{0}}\right| $

$d_{0}^{2}<1:$ $\Bbbk =-1$ ($d_{0}>0,\ b_{0}>0)$:$\ S>\left|
d_{0}b_{0}\right| ,\ S>\left| \frac{b_{0}}{d_{0}}\right| $


\begin{references}
\bibitem{ABH+96}  S. Aminneborg, I. Bengtsson, S. Holst, and P. Peldan,
Class. Quantum Grav. {\bf 13}, 2707 (1996).

\bibitem{SM97}  W.L. Smith and R.B. Mann, Phys. Rev. D {\bf 56}, 4942 (1997).

\bibitem{Lem98}  J.P.S. Lemos, Phys. Rev. D {\bf 57}, 4600 (1998).

\bibitem{DL01}  O.J.C. Dias and J.P.S. Lemos, {\it Magnetic strings in
anti-de Sitter General Relativity}, eprint, hep-th/0110202 (2001).

\bibitem{BLP97}  D.R. Brill, J. Louko, and P. Peldan, Phys. Rev. D {\bf 56},
3600 (1997).

\bibitem{Van97}  L. Vanzo, Phys. Rev. D {\bf 56}, 6475 (1997).

\bibitem{NVB02}  J.V. Narlikar, R.G. Vishwakarma, and G. Burbidge, {\it %
Interpretations of the Accelerating Universe}, eprint, astro-ph/0205064
(2002).

\bibitem{HP83}  S.W. Hawking and D.N. Page, Commun. Math. Phys. {\bf 87}, 57
(1983).

\bibitem{BTZ92}  M. Banados, C. Teitelboim, and J. Zanelli, Phys. Rev. Lett.
{\bf 69}, 1849 (1992).

\bibitem{BTZ94}  M. Banados, C. Teitelboim, and J. Zanelli, Phys. Rev. D
{\bf 49}, 975 (1994).

\bibitem{Pee00}  A.W. Peet, {\it TASI Lectures on Black Holes in String
Theory}, TASI-99, hep-th/0008241 (2000).

\bibitem{Isr77}  W. Israel, Phys. Rev. D {\bf 15}, 935 (1977).

\bibitem{FIU89}  V.P.\ Frolov, W. Israel, and\ W.G. Unruh, Phys. Rev. D {\bf %
39}, 1084 (1989).

\bibitem{HMV93}  S.J. Hughes, D.J. McManus, and M.A. Vandyck, Phys. Rev. D
{\bf 47}, 468 (1993).

\bibitem{SB98}  A.A. Sen and N. Banerjee, Astrophys. and Space Sci. {\bf 259}%
, 301 (1998).

\bibitem{MO54}  W. Magnus and F. Oberhettinger, {\it Formulas and Theorems
for the Functions of Mathematical Physics,} (Springer, New York, 1954).

\bibitem{Arf70}  G. Arfken, {\it Mathematical Methods for Physicists, }2nd Ed%
{\it , }(Academic Press, New York, 1970).

\bibitem{MTW73}  C.W. Misner, K.S. Thorne, and J.A. Wheeler, {\it Gravitation%
}, (W.H. Freeman, New York, 1973).

\bibitem{Wal84}  R.M. Wald, {\it General Relativity,} (University of Chicago
Press, Chicago, 1984).

\bibitem{BN90}  I. Brevik and H.B. Nielsen, Phys. Rev. D {\bf 41}, 1185
(1990).

\bibitem{Lem01}  J.P.S. Lemos, in {\it Astronomy and Astrophysics: Recent
Developments}, Proceedings of the 10th Portuguese Conference (World
Scientific, 2001).

\bibitem{WK80}  W.J. Wild and R.M. Kerns, Phys. Rev. D {\bf 21}, 332 (1980).
\end{references}
\end{document}